\documentclass[prm,notitlepage,twocolumn]{revtex4-2}
\usepackage{chemformula} 
\usepackage[T1]{fontenc} 
\usepackage{amsmath}
\usepackage{amssymb}
\usepackage{graphicx}
\usepackage{subcaption}
\usepackage{booktabs}
\usepackage{multirow}
\usepackage{tablefootnote}
\usepackage{float}
\usepackage{relsize}
\usepackage[version=3]{mhchem}
\usepackage{dcolumn}
\newcolumntype{.}{ D{.}{.}{-1} }
\usepackage{setspace}
\usepackage{ulem}
\usepackage{booktabs}
\usepackage{xcolor}
\usepackage{hyperref}
\usepackage{underscore}
\usepackage{comment}

\newcommand{\beq}{\begin{equation}}
\newcommand{\eeq}{\end{equation}}
\newcommand{\bea}{\begin{eqnarray}}
\newcommand{\eea}{\end{eqnarray}}

\newcommand{\benum}{\begin{enumerate}}
\newcommand{\eenum}{\end{enumerate}}
\newcommand{\bi}{\begin{itemize}}
\newcommand{\ei}{\end{itemize}}
\def\bfr{{\mathbf{r}}}

\def\sss{\scriptscriptstyle\rm}

\def\s{_{\sss S}}

\DeclareUnicodeCharacter{2212}{-}

\begin{document}

\author{H. Francisco}
\affiliation{Departamento de Qu\'imica, Universidad Aut\'onoma Metropolitana-Iztapalapa, Cd. de M\'exico 09340, M\'exico, and Quantum Theory Project, Dept.\ of Physics, University of Florida, Gainesville FL 32611, USA}
\email{francisco.hector@ufl.edu}

\author{Javier Carmona-Esp\'indola}
\affiliation{SECIHTI--Departamento de Qu\'imica, Universidad Aut\'onoma Metropolitana-Iztapalapa, Cd. de M\'exico 09340, M\'exico}

\author{A.C. Cancio}
\affiliation{Dept.\ of Physics and Astronomy, Ball State University, Muncie IN 47306, USA}

\author{Jos\'e L. G\'azquez}
\affiliation{ Departamento de Qu\'imica, Universidad Aut\'onoma Metropolitana-Iztapalapa, Cd. de M\'exico 09340, M\'exico}

\author{S.B. Trickey}
\affiliation{Quantum Theory Project and Center for Molecular Magnetic Quantum Materials,  Dept.\ of Physics,  University of Florida, Gainesville FL 32611, USA}
\email{trickey@ufl.edu}

\title{HOMO-LUMO Intervals from Modern Density Functionals \\ as Fundamental Gap Estimators}

\date{rev. 16 Sept. 2026; original 10 Aug. 2026}


\begin{abstract}
Use of the Kohn-Sham eigenvalue gap (highest occupied
molecular orbital to lowest unoccupied, HOMO-LUMO)
as a fundamental gap ($I-A$, $I=$ ionization
potential, $A=$ electron affinity) estimator is a long-standing challenge
for relatively simple exchange-correlation functionals,
e.g. meta-GGAs.  The LAK meta-GGA 
[T. Lebeda et al. Phys. Rev. Lett. \textbf{133}, 136402 (2024)] was developed to
emphasize certain constraints in ways intended to make the LAK KS gaps
close to the HSE06 hybrid functional
[A.V. Krudau et al. J.Chem. Phys. \textbf{125}, 224106 (2006)] gaps.  The LAK functional was
tested against a small atomization data set and a very small
bond length set, with no molecular HOMO-LUMO gap testing.
Here we test LAK against widely used, large reference data sets for
molecular heats of formation, bond lengths, harmonic frequencies,
$I$, and $A$ and compare
with r$^2$SCAN [J.W. Furness et al., J. Phys. Chem. Lett. \textbf{11}, 8208 (2020)] as well
as HSE06 and PBE0, a quite different hybrid.  On these tests LAK does not
give meaningfully better
geometries, heats of formation, HOMO-LUMO gaps, or 
DFT IP theorem satisfaction than r$^2$SCAN.  Compared against experimental
$I$ and $A$ values the same holds; LAK results resemble r$^2$SCAN
results, not HSE06.  Moreover, the HOMO and LUMO shifts
generated by the NCAPR GGA [J. Carmona-Esp\'indola et al., J. Chem. Phys.  \textbf{157}, 114109
  (2022)] as derivative discontinuity estimates, correct the
bare NCAPR HOMO and LUMO eigenvalues to substantially better
agreement with experiment for both fundamental gap and IP theorem
than do the meta-GGAs (LAK, r$^2$SCAN) or hybrids
(HSE, PBE0).
We discuss possible reasons why LAK behaves differently for molecules and solids.
\end{abstract}

\maketitle

\section{Introduction \label{sec:intro}}

Advances in the sophistication and quality of non-empirical,
constraint-based approximate exchange-correlation (XC) functionals
have been and continue to be foundational to the widespread use of 
Hohenberg-Kohn-Sham density functional theory (DFT) for computational
prediction of properties of materials and their molecular
constituents. An  underlying challenge
is the cost-accuracy balance associated with the unavoidable approximate 
XC functional. That compromise is determined by the intrinsic limits
and transferability of the specific XC approximation as well as user
priorities.

A long-standing example is the ``gap problem''. Simply put, to what
extent, if any, can an approximate XC functional that gives good
ground state energetics (e.g. atomization energies, bond lengths) be
used to estimate the fundamental gap
\beq E_g^{\mathrm{fund}} = I - A 
\label{eq:EgFund}
\eeq
(with $I$ the ionization
potential and  $A$ the electron  affinity),
from the Kohn-Sham eigenvalue gap 
\beq
E_g^{\mathrm{KS}} = \varepsilon_L - \varepsilon_H
\label{eq:EgKS}
\eeq
(highest occupied KS molecular
orbital eigenvalue to lowest unoccupied molecular orbital, HOMO-LUMO)?

There is contextual importance of the fundamental gap as well. For molecules, $I-A$ as in Eq. (\ref{eq:EgFund})
was identiﬁed by Parr and Pearson \cite{ParrPearson83} as the chemical hardness. It,
together with the chemical potential and the
Fukui function, constitutes the basis of conceptual density functional theory.
\cite{ParrWangBook,LiuBook}.

The formal properties of the KS eigenvalue gap are not at
issue. It has been known \cite {PerdewLevy83,ShamSchluter83} for
decades that the exact HKS functional has discontinuous derivative with respect to
particle number at integer values, $N$.  Almost all approximate functionals
at the meta-GGA or lower rungs of the Perdew-Schmidt ``Jacob's
ladder'' hierarchy of complexity \cite{PerdewSchmidt} violate
that. Again, the implications are much-discussed in literature 
so extensive that we cannot do a
review here. But see, for example,
Refs. \citenum{JohnsonYangDavidson2010,WYDD,YPSP2016} and references
therein.  The pertinent point is that for the exact $E_{xc}$
functional, the IP theorem \cite{IP_THEOREM1,PerdewLevy83,IP_THEOREM3}
\begin{equation}
  I = -\varepsilon_{\mathrm{H}}  
  \label{eq:IPthm}
\end{equation}
holds. (Remark: Sometimes this is misleadingly called the Koopmans condition for DFT, but the formal origins in DFT and Hartree-Fock are quite distinct.) 

For approximate functionals that are continuous at
integer $N$, the IP theorem is violated. The problem is the convexity
  of $E_x$ between $N-1$ and $N$ from an approximate functional instead of the required linearity
  \cite{IP_THEOREM1}. One remedy for that is to correct an
  existing exchange functional
  with a term that yields approximately the required linearity.  Extensive
  effort in that regard has been made by Yang and co-workers 
  \cite{ZhengCohenEtAl2011,ZZY2013,ZZLY2015,Zhang2018,LZSY2018,YZY2020,MCY2020}.  A challenging but potentially worthwhile alternative
pursuit then is to construct a constraint-based non-empirical
approximate functional which substantially reduces the 
convexity error to the point of delivering KS gaps that are
useful approximations to the fundamental gap while preserving or even bettering the performance of
 an antecedent functional or functionals on basic thermochemical properties.

One route for that is via the KS-orbital dependence
that appears in meta-generalized-gradient approximation (meta-GGA) XC functionals but not in GGAs. 
For the meta-GGA functionals of interest here, the orbital-dependence
comes from use of the non-interacting 
(Kohn-Sham, KS) kinetic energy density~\cite{BeckeEdgecombe,SunEtAl2013},
\beq
\tau\s  :=  \frac{1}{2}\sum_i f_i |\nabla \varphi_i(\bfr)|^2
\label{eq:tausdefn}
\eeq
with KS orbitals $\varphi_i$ and occupation numbers $f_i$. (We use
Hartree atomic units throughout unless noted to the contrary.)
The provocative clue is that so-called hybrid functionals, in which there is an orbital-dependent
contribution from explicit single-determinant exchange, typically
do better on the gap problem.  

Recently a meta-GGA functional (LAK for Lebeda, Aschebrock, K\"ummel)
\cite{lak1,lak2} has been constructed with the design goal of
improving substantively upon the SCAN 
\cite{SCAN,SCANNature} functional regarding the gap problem
while preserving its performance on basic ground-state thermochemical
properties. The target gap behavior was that delivered by the HSE06 hybrid \cite{HSE06}.
The LAK construction  is predicated on  
the observation that the gradient expansion for meta-GGAs contains a previously 
unexplored degree of freedom that allows shifting the relative weight
between density-gradient and kinetic-energy-density contributions.
Ref.\ \citenum{lak1} asserts that LAK gives remarkably accurate
descriptions of electronic bonds and band gaps, positioning it as a
potentially high-performing exchange–correlation approximation obtained 
by exploiting previously un-utilized gradient expansion properties.

In Ref.\ \citenum{lak1} 
there is somewhat more emphasis on outcomes for periodic solids
than for isolated molecules.  In  comparison with fairly
common molecular testing (e.g. Ref. \citenum{MVSrevisited})
the molecular data sets used in Ref. \citenum{lak1} were small.
No results were given for molecular KS 
gaps.  Here we report the results of detailed,
large-data-set assessment of the LAK functional for molecular
thermochemistry, gaps, and IP theorem.  The IP theorem
and fundamental gap comparisons are both with respect to satisfaction
between eigenvalues and total energy differences to calculate $I-A$
and with respect to experimental values for $I$ and $A$ on an extensive
data set.  In the next section we
define the metrics used and summarize the technical details
and choices.  Section 3 is devoted to the numerical results and
analysis of them.  Section 4 provides a brief discussion of observations
and conclusions.

\section{Methods and Metrics \label{sec:MM}}

To provide performance comparison and context, we selected a suite of
several functionals.  First, for a baseline reference from semi-local
functionals, we included PBE \cite{PBE}, the most widely used GGA.  It
exhibits the gap problem quite clearly. For comparison at the GGA
level of complexity, we tested the NCAPR XC functional
\cite{JCP157}. It was constructed to have very different behavior from
PBE. That behavior enables the derivative discontinuity to be
estimated and applied, post-scf, to the NCAPR KS eigenvalues. For
meta-GGAs, in
addition to LAK we included r$^2$SCAN \cite{FurnessEtAl2020}, a
refined version of SCAN. Currently it is the most competitive
non-empirical meta-GGA.  Regarding hybrids, in addition to HSE06 we
included PBE0 \cite{PBE01,PBE02}. It consists of 75\% PBE exchange,
25\% single-determinant X, and 100\% PBE correlation, quite different
from the screened hybridization of HSE06.

Following the same protocol used in previous studies
\cite{MVSrevisited}, the molecular calculations were performed using
NWChem 7.2.3 \cite{NW} with the def2-TZVPP basis set and \textit{xfine}
integration grids. Note, in particular, that the meta-GGA and
  hybrid calculations were done in the simple generalized Kohn-Sham framework, i.e. with an orbital-dependent exchange-correlation potential.\cite{YPSP2016}

The performance of the suite of XC functionals was evaluated
on the widely used G3X/99 \cite{G31,G32}, T96-R, and T82-F
\cite{DF1,DF2} molecular test sets.  For the G3X/99 molecules, the
corresponding HOMO and LUMO energies were calculated as well as the
conventional heats of formation. (Unless otherwise stated or where it
is obviously not the case, all total-energy and electronic-structure
calculations reported in this Section were done using the fixed
reference geometries prescribed for heat of formation calculation
from the original G3X/99 dataset.)
 The T96-R set was used to determine
optimized bond lengths, while the T82-F set provided harmonic
vibrational frequencies. We remind the reader that these standard benchmark
  data sets 
  are comprised of comparatively small molecules (mostly ten or fewer atoms).
  Performance with respect to those benchmarks therefore gives no
  clue as to possible evolution of behavior with respect to cluster size.

Motivated by the gap and IP-theorem results for those widely used data
sets, we proceeded to comparison with experimental $I$ and $A$ values
using the 83 molecule data set compiled earlier by some of us
\cite{JCP157}. Note that the 38 $A < 0$  experimental values
and associated computed neutral molecule geometries originated in Ref. \citenum{Zhang2018}. For brevity, we refer to the full data-set as ``IA-83''.

Difficulties with implementations of HSE06 are well-known
\cite{HSE06difficulties}. At the outset of this study we found that
the native NWChem HSE06 implementation is faulty.  Following a
detailed analysis, we ascertained that the Libxc implementation of
HSE06 is correct \cite{LIBXC}.  We have used it throughout.

Regarding LAK, Libxc does not provide the correlation part.  To
surmount this difficulty, we utilized a development version of Libxc
privately shared by T. Lebeda.

In addition to the customary thermochemical tests, we calculated the
ionization potential (IP), electron affinity (EA), and fundamental gap
for all the molecules with each of the functionals. In terms of total
energy differences those are
\begin{equation}
  I = E_{N-1} - E_{N} \;,
 \label{eq:IPdefn}
\end{equation}
and
\begin{equation}
  A = E_{N} - E_{N+1} \;,
   \label{eq:EAdefn}
\end{equation}
with $E_{N}$, $E_{N-1}$, and $E_{N+1}$ being the neutral, cation, and
anion system total energies respectively.  Anion calculations can
require special attention.  We discuss that when pertinent in what
follows.  Remark: For isolated molecules, $\Delta$-SCF
  calculations provide a direct route to ionization potentials,
  electron affinities, and
fundamental gaps through total-energy differences. As discussed below,
$\Delta$-SCF calculations were used in this study.  Crucially,
however, such calculations are considerably less straightforward for
periodic solids, which is one reason why eigenvalue-based gap
estimates remain of significant practical interest in that context.

An obvious  metric of the degree to which a functional
satisfies the IP-theorem is
\beq
\Delta_I := \varepsilon_{\mathrm{HOMO}} + I  \;
\label{eq:DeltaIdefn}
\eeq
It follows from Eq. (\ref{eq:IPthm}) that $\Delta_I=0$
for the exact functional and a complete basis, so
deviations from $\Delta_I = 0$ quantify the extent of IP-theorem
noncompliance.  Notice that this is written as an internal consistency
check on a functional: both $I$ and $\varepsilon_{\mathrm{HOMO}}$
are calculated quantities. Clearly $\Delta_I$ also can be
evaluated against experimental $I$ data. We return to this
in the context of the IA-83 dataset. 

A related but more demanding measure of functional behavior is
the discrepancy between the Kohn-Sham gap and the fundamental gap,
to wit
\begin{equation}
\Delta E_g = E_g^{\mathrm{KS}} - E_g^{\mathrm{fund}} \;.
\label{eq:DeltaEgdefn}
\end{equation} 
 The ``gap problem'' is solved for $\Delta E_g = 0$. Thus, $\Delta E_g$ 
 is a measure of the extent to which a given approximate functional
reproduces or mimics the derivative discontinuity.

\section{Results \label{sec:results}} 
\subsection{Commonplace Thermochemical Benchmarks \label{subsec:one}}

Table \ref{Table:mol_v1} provides a summary
comparison of LAK results with those from the PBE \cite{PBE} and NCAPR \cite{JCP157} GGAs,
the r$^2$SCAN \cite{FurnessEtAl2020} meta-GGA,
and the HSE06 \cite{HSE06} and PBE0 \cite{PBE01,PBE02} hybrid functionals.
(Molecule-by-molecule detailed data are given in Tables S1-S3 of the
Supplementary Material.) Overall, the LAK and r$^2$SCAN results are
very similar on these standard molecular benchmarks and exhibit
performance characteristic of modern semi-local functionals, not those of 
hybrid approximations. For heats of formation, HSE06
has a much smaller mean error magnitude and minusculely smaller
mean absolute error than both LAK and r$^{2}$SCAN. PBE0 yields slightly higher values than the meta-GGA functionals and the range-separated hybrid HSE06, whereas NCAPR and especially PBE show substantially larger deviations.
For equilibrium bond lengths,  all hybrid and meta-GGA functionals perform comparably, while NCAPR and PBE exhibit noticeably larger errors.  At least for
those two datasets, the inclusion of single-determinant exchange
does yield an improved description of thermochemical properties.
In contrast, for harmonic vibrational
frequencies, HSE06 displays larger errors than both LAK and
r$^{2}$SCAN.  LAK is slightly worse on frequency MAD and substantially
worse on MAD than r$^{2}$SCAN. 

\begin{table*}[t]
\centering
\caption{Results for molecular test sets G3X/99, T96-R, and T-82F respectively for the LAK exchange-correlation functional compared with r$^{2}$SCAN, HSE06, PBE0, NCAPR, and PBE. ME= mean error, MAD=mean absolute
  deviation. Heat of formation errors in $kcal/mol$, bond length errors in $\AA$, and frequency errors in $cm^{-1}$.}
\renewcommand{\tabcolsep}{12pt}
\begin{tabular}{@{}llcccccc@{}}
\toprule
     &      &LAK & r$^2$SCAN & HSE06 & PBE0 & NCAPR &PBE\\
    
\midrule
\hline
  \multirow{3}{*}{Heats of Formation} & ME  &-2.204&-3.145&-0.633&-3.330&-5.688&-20.878\\
                                      & MAD &4.618&4.488&4.394&5.728&7.353&21.385\\
                                      &     &        &        &\\
  \multirow{3}{*}{Bonds}              & ME  &0.006&0.005&0.000&-0.001&0.025&0.018\\
                                      & MAD &0.010&0.010&0.010&0.010&0.025&0.018\\
                                      &     &        &         &\\
  \multirow{3}{*}{Frequencies}        & ME  &20.36&11.34&31.98&34.33&-47.00&-33.78\\
                                      & MAD &32.82&30.90&43.44&44.51&52.45&43.61\\
\hline 
 \bottomrule
\end{tabular}
\label{Table:mol_v1}
\end{table*}

To gain detailed insight on comparative band-gap performance in
molecules, we returned to the 223 molecule G3X/99 dataset and 
evaluated key electronic structure properties using the six XC
suite as before. For each molecule and all the functionals,  
total energies and Kohn-Sham eigenvalues (HOMO and LUMO) were obtained
for both the neutral systems ($N$ electrons) and the charged systems
with ($N-1$) and ($N+1$) electrons.  Keep in mind, as stated
already, these calculations were at G3X/99 prescribed geometries.
Numerically (and not unexpectedly) the $E_{N+1}$ calculations
generally were more demanding than the neutrals or cations. In a subset
of systems, a relatively larger number of
self-consistent field (SCF) iterations was required to achieve
convergence. However, the same convergence tolerances ultimately
were satisfied for all systems and for all six functionals.

Table \ref{Table:g3_metrics} presents a statistical summary regarding the
two gap indicators for the
six functionals and the 223 molecules in the G3X/99 dataset. (Detailed molecule-by-molecule results are in the Supplementary Material, Tables S4--S11.) 
 For all the
functionals, the mean error for $\Delta_I$ is positive, reflecting a
systematic deviation from the DFT IP theorem. For LAK and
r$^2$SCAN the performance is essentially 
indistinguishable. Consistent with  being GGAs, NCAPR and PBE show the largest deviations. (Note that the NCAPR eigenvalue results are `''bare'', that is,  \textit{without} the
NCAPR derivative-discontinuity shifts.  We provide results on that in the discussion below of the IA-83 dataset study.)
PBE0 exhibits the smallest deviations for both $\Delta_I$ and $\Delta E_g$, followed by HSE06. Given the use of HSE06  $\Delta E_g$ values as gap targets, this perhaps is
an interesting finding.  

\begin{table*}[t]
\centering
\caption{Global statistical analysis over the full set of 223 molecules for the LAK, r$^2$SCAN,  HSE06, PBE0, NCAPR, and PBE functionals. The table reports the mean error (ME), mean absolute deviation (MAD),
and Spread (defined as the difference between the largest and smallest signed errors) for IP theorem deviation $\Delta_I$ and for $\Delta E_g$ (the difference between fundamental and KS gaps). All values are in eV.}
\renewcommand{\tabcolsep}{12pt}
\begin{tabular}{@{}lcccccccc@{}}
\toprule
DFA    & ME($\Delta_I$ ) & MAD($\Delta_I$) & Spread($\Delta_I$ ) & ME($\Delta E_g$) & MAD($\Delta E_g$) & Spread($\Delta E_g$) \\
\midrule
\hline
LAK       & 3.598 & 3.598 & 7.022 & -6.370 & 6.370 & 13.663 \\
r$^2$SCAN & 3.610 & 3.610 & 7.198 & -6.525 & 6.525 & 14.154 \\
HSE06     & 3.013 & 3.013 & 5.783 & -5.731 & 5.731 & 12.068 \\
PBE0      & 2.614 & 2.614 & 5.741 & -4.974 & 4.974 & 11.917 \\
NCAPR     & 3.814 & 3.814 & 7.123 & -6.978 & 6.978 & 14.741 \\
PBE       & 3.831 & 3.831 & 7.147 & -7.133 & 7.133 & 14.838 \\
\hline 
 \bottomrule
\end{tabular}
\label{Table:g3_metrics}
\end{table*}

The $\Delta E_g$ (recall Eq. \ref{eq:DeltaEgdefn}) behavior
reflects the known fundamental gap
underestimation characteristic of semi-local
functionals. All six 
functionals exhibit negative mean errors, which reflects an
underestimation of $E_g^{\mathrm{fund}}$ by the KS gap,
$E_g^{\mathrm{KS}}$. Interestingly, this holds even for the
hybrids.  

The LAK MAD and spread values for $\Delta E_g$ are only slightly
better than those for r$^2$SCAN. In particular, the spread decreases
from 14.15 to 13.66 eV, indicating a modest reduction in the overall
error range. Although PBE0 exhibits the smallest overall errors, LAK
still provides a modest improvement over r$^2$SCAN. PBE and NCAPR
(again with bare eigenvalues) have considerably larger errors, as would
be expected from GGAs.

These results, taken together, indicate that LAK is not significantly
superior to  r$^2$SCAN for molecules.  In particular,
it is not meaningfully better at IP-theorem compliance. 

There is a technical issue to consider. The $\Delta_I$ and $\Delta
E_g$ results just presented were obtained using the fixed reference
geometries from the G3X/99 dataset.  That allows for a direct
comparison of the electronic performance of the different functionals
in the context of the standard G3X/99, T-96R, T-82F thermochemistry
assessment.  But bench-marking at fixed geometry runs the risk of suppressing
significantly different geometry relaxation effects for functionals constructed and parametrized on
substantively different constraints and principles. Compared,
for example to r$^2$SCAN, that is precisely the case with LAK.  

To probe that issue, we did additional calculations on the G3X/99
molecules in which the geometry of the neutral system was optimized
self-consistently with each XC functional.  That
neutral geometry then was employed in the subsequent total-energy and
eigenvalue calculations for both the neutral and charged states. The
same basis set and numerical settings were used as in the
fixed-geometry calculations. Detailed results are in the Supplementary
Material Tables S12--S19.

Table \ref{Table:g3_metrics_opt} makes clear that the use of optimized
neutral geometries associated with a specified functional leads to
only minor quantitative changes, relative to the fixed-geometry
results, in the global error metrics. In particular, the same ordering
and the same relative
performance of the different functionals occurs. LAK and
r$^{2}$SCAN still exhibit nearly identical IP-theorem deviations,
while LAK retains a small, almost negligible but systematic,
improvement over r$^{2}$SCAN in the description of the fundamental
gap. Neither one is really competitive with HSE06 for either
$\Delta_I$ or $\Delta E_g$. PBE remains the least accurate functional
for both metrics.  Clearly the gap trends found at fixed (G3X/99)
geometries are consistent with those found with relaxed geometries:
LAK does not mimic HSE06, despite having been designed with that
as an objective. 

\begin{table*}[t]
\centering
\caption{As in Table \ref{Table:g3_metrics} for the 223 G3X/99 molecules, \textit{but} using neutral species geometries optimized with each functional.}
\renewcommand{\tabcolsep}{12pt}
\begin{tabular}{@{}lcccccccc@{}}
\toprule
DFA    & ME($\Delta_I$ ) & MAD($\Delta_I$) & Spread($\Delta_I$ ) & ME($\Delta E_g$) & MAD($\Delta E_g$) & spread($\Delta E_g$)  \\
\midrule
\hline
LAK    & 3.611 & 3.611 & 7.051 & -6.395 & 6.395 & 13.698 \\
r$^2$SCAN & 3.626 & 3.626 & 7.227 & -6.545 & 6.545 & 14.192 \\
HSE06  & 3.014 & 3.014 & 5.814 & -5.736 & 5.736 & 12.452 \\
PBE0   & 2.611 & 2.611 & 5.770 & -4.962 & 4.962 & 12.306 \\
NCAPR  & 3.784 & 3.784 & 7.141 & -6.958 & 6.958 & 14.726 \\
PBE    & 3.824 & 3.824 & 7.160 & -7.123 & 7.123 & 14.823 \\
\hline 
 \bottomrule
\end{tabular}
\label{Table:g3_metrics_opt}
\end{table*}

\subsection{Comparison with Experimental $I$ and $A$ Values  \label{subsec:two}}

The foregoing accuracy tests are internal in the sense that they are
measures of satisfaction of rigorously provable equalities in DFT (and quantum
mechanics more generally).  Self-evidently such consistency tests are
incomplete in that they provide no direct insight into experimental
fidelity.  Thus we turn to 
comparison with measured $I$ and $A$ values. Though there are data for
fewer molecules, the IA-83-molecule dataset \cite{JCP157} is equipped with 
experimental values.  The associated molecular geometries  for $A>0$ systems
are neutral system results 
from B3LYP calculations with 6-31G*, or 6-31G(2df,p) basis sets.
The geometries for the 38 systems with $A<0$ are from B3LYP
calculations with 6-311+G** basis.\cite{Zhang2018} Those neutral system geometries
were used for the anions and cations as well. 
See Ref. \citenum{JCP157} for details.

Compilation of IA-83 was motivated
by work on the NCAPR GGA functional \cite{JCP157}.  We mentioned already that it
is a GGA that produces a
corrected KS gap by way of analysis of its exchange potential.
The NCAPR X potential goes to a
positive constant as $-c/r$, $c \approx 0.3$ for a molecular system.  Shifting
down to the physically required zero provides a way to calculate an approximate
derivative discontinuity shift to the NCAPR eigenvalues. Shifting
does not, of course, change the NCAPR total energy.  We include
\textit{shifted NCAPR} eigenvalue results in what follows.

A pertinent methodological difference relative to the procedure used
in Ref.\ \citenum{JCP157} is the treatment of anions. In the interest
of computational efficiency, the calculations in that work started the
SCF cycle with basis set coefficients projected from a smaller basis
(cc-pVTZ).  In the present work, with a far faster machine available,
it was natural to start directly with the full aug-cc-pVTZ
basis. Doing that uncovered an unexpected aspect of the previous
technique. Many of the anions are so delicately bound that starting with
the projected cc-pvTZ state leads to a converged state that is higher
in energy than the converged state (with the same tolerances) found
from using the aug-cc-pVTZ basis from the start. \textit{Thus we
  caution the reader.}  We have no evidence that the anion energies
reported here are not minimal.  They differ, therefore, from those
reported in Ref.\ \citenum{JCP157} because of the unexpected effect of
the difference in techniques.  Moreover, the whole matter of anions
treated with ground-state DFT and finite basis sets is fraught with
difficulties \cite{NR-SBTNegIons,PeachEtalIons2015}.  For all these
reasons, we are not as confident in the results for $A < 0$ obtained
as total energy differences as for
those with $A > 0$.

\begin{table*}[t]
\centering
\caption{Summary of statistical errors (ME, MAD, and Spread, in eV) for ionization potentials $I$, electron affinities $A$, and fundamental gaps $I-A$
from total energy differences, and the corresponding quantities obtained from Kohn--Sham
eigenvalues. Errors are relative to the experimental $I$ and $A$ values
tabulated in IA-83. ``Shifted NCAPR'' denotes eigenvalues shifted
by the derivative discontinuity estimate given by NCAPR; see text.}
\renewcommand{\tabcolsep}{8pt}
\begin{tabular}{@{}llccccccc@{}}
\toprule
Property & & PBE & r$^2$SCAN & HSE06 & PBE0 & NCAPR & LAK & Shifted--NCAPR \\
\midrule

\multirow{4}{*}{Total-energy IP}
 & ME     & 0.010 & 0.049 & 0.120 & 0.126 & 0.060 & 0.020 & --- \\
 & MAD    & 0.351 & 0.347 & 0.317 & 0.313 & 0.344 & 0.350 & --- \\
 & Spread & 3.175 & 3.247 & 3.503 & 3.501 & 3.127 & 3.228 & --- \\
&&&&&&&&\\

\multirow{4}{*}{Total-energy EA}
 & ME     & 0.304 & 0.183 & 0.225 & 0.203 & 0.285 & 0.212 & --- \\
 & MAD    & 0.620 & 0.586 & 0.603 & 0.585 & 0.582 & 0.608 & --- \\
 & Spread & 5.149 & 5.265 & 5.239 & 5.233 & 5.122 & 5.303 & --- \\
&&&&&&&&\\

\multirow{4}{*}{Fundamental Gap}
 & ME     & -0.295 & -0.134 & -0.106 & -0.077 & -0.225 & -0.191 & --- \\
 & MAD    & 0.881 & 0.854 & 0.847 & 0.821 & 0.817 & 0.872 & --- \\
 & Spread & 6.963 & 6.108 & 6.154 & 6.144 & 5.868 & 6.081 & --- \\
&&&&&&&&\\

\multirow{4}{*}{HOMO IP}
 & ME     & -3.660 & -3.388 & -2.744 & -2.345 & -3.620 & -3.379 & 0.270 \\
 & MAD    & 3.660 & 3.388 & 2.744 & 2.345 & 3.620 & 3.379 & 0.581 \\
 & Spread & 3.807 & 3.450 & 2.758 & 2.754 & 3.799 & 3.426 & 2.943 \\
&&&&&&&&\\

\multirow{4}{*}{LUMO EA}
 & ME     & 2.900 & 2.532 & 2.211 & 1.842 & 2.844 & 2.414 & 0.466 \\
 & MAD    & 2.900 & 2.532 & 2.211 & 1.852 & 2.844 & 2.417 & 0.651 \\
 & Spread & 5.381 & 4.654 & 4.338 & 4.404 & 4.091 & 4.743 & 3.832 \\
&&&&&&&&\\

\multirow{4}{*}{KS Gap}
 & ME     & -6.560 & -5.921 & -4.955 & -4.187 & -6.464 & -5.793 & -0.196 \\
 & MAD    & 6.560 & 5.921 & 4.955 & 4.187 & 6.464 & 5.793 & 0.926 \\
 & Spread & 7.077 & 5.829 & 4.884 & 4.824 & 6.940 & 5.675 & 5.191 \\
\hline
\bottomrule
\end{tabular}
\label{Table:summary}
\end{table*}

The Supplementary Material Tables S20-S25 provide  
detailed  $I$, $A$, and $I-A$ results for all 83 molecules as
calculated from total energy differences (Eqs. \ref{eq:IPdefn},
\ref{eq:EAdefn}) and from Kohn--Sham eigenvalues from PBE, r$^2$SCAN,
  HSE06, PBE0, NCAPR, and LAK. Table \ref{Table:summary} summarizes
  those results.  That provides direct comparison among
  conventional GGAs, meta-GGAs, global hybrids, and an approximate
  derivative-discontinuity correction applied to the Kohn--Sham
  eigenvalues.

Consider first the ``gap problem'' that motivated this study.  The
last three sections of that Table provide data to address it.  The Table section
labeled ``HOMO-IP'' is, in essence, an evaluation of $\Delta_I$
Eq. (\ref{eq:DeltaIdefn}) with the experimental IP as the reference.
Similarly, the Table section labeled ``KS Gap'' is, in fact, an
evaluation of $\Delta E_g$ Eq. (\ref{eq:DeltaEgdefn}) with respect to
experimental $I - A$ values.

Clearly the most pronounced failure is for HOMO eigenvalues with
respect to the IP theorem, the $\Delta_I$ test.  Ionization potentials
estimated by HOMO energies are underestimated by $\approx 3$--$4$ eV
for all the semi-local functionals, including LAK.  Thus, LAK does not
emulate HSE06 for $\Delta_I$ with $I$ from experiment. The HSE06 MAD
is 2.744 eV (and PBE0 is better at 2.345 eV) versus LAK of 3.379 eV. The LAK value is not meaningfully
better than that from r$^2$SCAN (3.388 eV) or even from PBE (3.660
eV). The same appraisal holds for the Spreads. 

Since LAK was designed specifically to reproduce HSE06 Kohn-Sham gaps,
it makes sense to look at that  quantity with extra care.  
The test of $\Delta E_g$  against experiment provided by the last section
of Table \ref{Table:summary} follows the same pattern as the IP theorem
tests.  Note, first, that
HSE06 did not really solve the ``gap problem''.
Its MAD for $\Delta E_g$ is 4.955 eV. 
However, if we take HSE06 as the target, LAK, at MAD of 5.793 eV is 17\% worse,  
not substantively better on $\Delta E_g$ than r$^2$SCAN (MAD = 5.921 eV). Their spreads are
similar too.

A more informative gauge of LAK reproduction of HSE06 Kohn-Sham gaps
may be to compare functional differences with the HSE06 results
directly. Measured on the G3X/99 test set, the spread of differences
of LAK with respect to HSE06 is 2.65~eV. This is considerably smaller
than the spread of differences with experiment (5.68~eV), and better
than r$^2$SCAN spread with respect to LAK of 3.32~eV.  So LAK KS gap
performance does
correlate with HSE06 performance with modest success.  But LAK gaps are
systematically smaller than HSE06 gaps.  The average LAK gap is
0.84~eV smaller than that from HSE06 with a standard deviation of
0.45~eV. Following standard statistical analysis, the likely error in
the LAK gap measurement over a data set the size of the G3X/99 is a
mere 0.03~eV, far too small to explain LAK disagreement with HSE06
through random noise.

The one exception to the general $\Delta_I$ behavior found in Table
\ref{Table:summary} is provided by the shifted NCAPR eigenvalues
(recall above about how the shift is calculated).  Those eigenvalues
reduce the ``HOMO IP'' MAD relative to experiment to $\approx 0.6$ eV
and the $\Delta E_g $ (``KS gap'') error to about an eV.
Something similar occurs for the electron affinity. The calculated
shifts lower the NCAPR MAD to 0.65 eV compared to the  bare NCAPR eigenvalue
MAD of 2.84 eV (essentially the same as PBE, the other GGA tested).
Interestingly, even PBE0 remains substantially less accurate in this
context than the shifted-NCAPR results. That finding reinforces the importance
of correcting the derivative discontinuity effect upon eigenvalues.
Such correction is not achieved by hybridization.

In contrast to plain eigenvalue differences, the fundamental gaps
calculated from total energy differences are in much better agreement
with experiment than the KS gaps, precisely as expected from both
theory and experience.  Those differences are remarkably similar
for both the semi-local and hybrid functionals.  

LAK performance in this context is quite similar to that of other
semi-local functionals. It gives rather accurate results for total energy
differences, while inaccurate as to the HOMO eigenvalue. This
is, of course, consistent with the observation that the problem is
characteristic of the particular rung of the Perdew-Schmidt ladder
(i.e. intrinsic to semi-local approximations), not of a specific
functional on that rung.  Again, the exception is NCAPR, which was
built deliberately to a very different set of constraints than the
usual GGA or meta-GGA.  

\section{Discussion and Concluding remarks \label{sec:conclusion}}

We are left with the interesting observation that, although LAK was
designed to improve KS gap predictions, those design principles appear
to be substantially more effective for crystalline solids than for
isolated molecules.  For the ground state, this is a peculiar but not
completely unfamiliar dilemma.  A similar difference exists with
the OFR2 de-orbitalization of r$^2$SCAN \cite{OFR2}.  On standard test
sets it favors solids to the detriment of molecules. The r$^2$SCAN-L
de-orbitalization, which violates the gradient expansion, is better
balanced between molecules and solids \cite{OFR2,r2SCANL}. At the less
flexible, GGA-level of refinement, original PBE somewhat favors
molecules.  PBEsol was devised to cope with that \cite{PBE,PBEsol}.
PBEmol \cite{PBEmol} goes the other way by using self-interaction
cancellation for the H atom as a constraint.
A common thread is that LAK, OFR2, and PBEsol all make detailed,
stringent use of the exchange gradient expansion.

But those are ground state trends.  For the gap problem, matters are
less clear.  A key assumption in LAK development is that a particular
kind of gradient expansion compliance improves KS gaps as estimators
of fundamental gaps.  The present results show that not to be true for
isolated molecules. As to other meta-GGAs and related GGAs, Table III
of Ref. \citenum{SRPP26} shows that the crystalline band gap MAD and
spread for r$^2$SCAN, r$^2$SCAN-L (the original,
gradient-expansion-violating de-orbitalization of r$^2$SCAN), and OFR2
(the gradient compliant de-orbitalization) are (1.20, 4.32), (1.38,
4.72), and (1.64, 4.99) eV respectively.  That pattern is consistent
with the notion that gradient expansion compliance can be troublesome.

However, there also is molecular data for PBE, PBEsol, and
PBEmol.\cite{PBEmol} The MADs for the IP for PBEsol, PBE, revPBE, and
PBEmol are respectively 2.63, 3.47, 3.0, and 3.97 kcal/mol. PBEsol is
the gradient-expansion compliant one, PBEmol is the farthest from
compliance.  Relative to PBE, PBEmol improves molecular heat of
formation performance dramatically, but at the cost of worsened
IP-theorem compliance.  (revPBE has a different functional form than
the other three, allowing competing design issues to come into play.)

The evidence to date thus suggests that there may not be a
straightforward relationship between gradient expansion compliance and
IP theorem satisfaction, with observed trends affected by competing
factors such as functional form.  It is plausible that solids (or at
least solids in the datasets used in testing) have more significant
spatial regions with density that resembles gradient expansion
behavior than do molecules.  This suggests that design strategies
focused on band-gap predictions in periodic systems may need to be
augmented or modified to provide results of counterpart quality for
molecules.  Investigation of these issues clearly is beyond the scope
of the present work.

The comparison presented here also indicates that improving the
description of molecular K-S gaps to ameliorate the gap problem
requires more than reproducing hybrid-like band gaps in solids. Within
the present benchmark, LAK is competitive with other modern semi-local
functionals. But despite its having been designed and constructed to 
achieve near-hybrid-level performance for
molecular ionization potentials, electron affinities, and  KS gaps,
it does not do so in general.

\section*{Supplementary Material}
The Supplementary Material contains:
\begin{itemize}
  \item detailed thermochemical results,
   \item molecule-by-molecule electronic properties for the G3X/99 test set (fix-geometry),
    \item optimized-geometry benchmark tables for the G3X/99 test set,
     \item electronic properties for the IA-83 test set.
\end{itemize}

\begin{acknowledgments}

We present this paper in memory of Axel Becke and as a modest tribute
in honor of his multiple contributions to DFT.  Perhaps the simplest
tribute is best: Many of the exchange-correlation functionals studied
and compared here incorporate his innovations in critical ways.
  
We thank T. Lebeda for providing a private version of the Libxc 7.0.0
tarball that includes the correlation part of the LAK functional.

HF was supported by SECIHTI-Mexico under the Postdoctoral Stays for Mexico Program 2025 (EPM 2025).

The work of SBT was supported as part of the Center for Molecular
Magnetic Quantum Materials, an Energy Frontier Research Center funded
by the U.S. Department of Energy, Office of Science, Basic Energy
Sciences under Award No.\ \mbox{DE-SC0019330}.

\end{acknowledgments}

\section*{Author Declarations}

\textbf{Conflicts of Interest} 

The authors have no conflicts to disclose.


\newpage

\end{document}